\documentclass[letterpaper,english,aps,prb,reprint,amsmath,amssymb,onecolumn]{revtex4-2}
\usepackage[T1]{fontenc}
\usepackage[latin9]{inputenc}
\usepackage{babel}
\usepackage{amsmath}
\usepackage{amssymb}
\PassOptionsToPackage{normalem}{ulem}
\usepackage{ulem}
\usepackage{microtype}
\usepackage[unicode=true,
 bookmarks=true,bookmarksnumbered=false,bookmarksopen=true,bookmarksopenlevel=1,
 breaklinks=false,pdfborder={0 0 1},backref=false,colorlinks=false]
 {hyperref}

\makeatletter

\pdfpageheight\paperheight
\pdfpagewidth\paperwidth

\usepackage{braket}
\usepackage{mathtools}

\makeatother

\begin{document}
\preprint{APS/123-QED}
\title{Universal secular form of non-secular master equation}
\author{Le Tuan Anh Ho}
\email{chmhlta@nus.edu.sg}

\affiliation{Department of Chemistry, National University of Singapore, 3 Science Drive 3 Singapore 117543}
\affiliation{Theory of Nanomaterials Group, Katholieke Universiteit Leuven, Celestijnenlaan 200F, B-3001 Leuven, Belgium}
\author{Liviu F. Chibotaru}
\email{liviu.chibotaru@kuleuven.be }

\affiliation{Theory of Nanomaterials Group, Katholieke Universiteit Leuven, Celestijnenlaan 200F, B-3001 Leuven, Belgium}
\date{\today}
\begin{abstract}
We develop a more accurate secular form for the Redfield non-secular master equation, which is insensitive to the choice of the basis. This completely solves the ambiguity on which basis should be used when the secular approximation of the Redfield master equation is applied.

\end{abstract}
\maketitle

\section{Introduction}

\global\long\def\hmt{\mathcal{H}}%
\global\long\def\vt#1{\vec{#1}}%

Real physical systems are always in interaction with its surroundings, and spin system is not an exception. In non-equilibrium state, a spin in contact with a reservoir will undergo a process of relaxation where it exchanges energy with environment of which quantized particles of the lattice vibrations, phonons, is one of the most important contributions at high temperature. Understanding this spin-phonon relaxation process is thus crucial for many fundamental problems and applications such as quantum coherence/decoherence in spin-based qubit \citep{Zurek2003,Stamp2009,Golovach2004,Vincent2012}, implementation of spintronics devices \citep{Zutic2004,Bogani2008,Mannini2010,Ardavan2007}, or high-density information storage/quantum computing using single-molecule/single-atom magnets \citep{sieklucka2016molecular,Leuenberger2001,Sessoli1993,Zhang2015a,Dunbar2012}. Of all the possible scenarios, a spin in weakly coupling with the phonon bath is often occurred. In this scenario, the spin dynamics can be described by the Redfield equation, which makes use of Born approximation of weakly coupling and Markov approximation of short memory \citep{Blum1996,May2011}. 

Formulated in the operators form, the Redfield master equation can have different matrix forms depending on the basis. In molecular magnetism generally and single-molecule/single atom magnets particularly, existing works so far can be divided into two groups: one uses the localized (natural) basis and the other uses eigenstates (diagonal) basis \citep{Leuenberger2000,Mannini2010,Garanin2011,Garanin1997,Villain1994,Gomez-Coca2014,Hartmann-Boutron1996a,Luis1998}. In principle, the matrix forms of the Redfield master equation in these two bases are equivalent \citep{Garanin2011}. However, due to its complexity in practice, most of the theoretical works only used the secular approximation of the matrix form of the master equation, either in localized basis or eigenstates basis. Unfortunately, these secular master equations are not equivalent and so are its solutions. As a consequence, the results and/or interpretations of involved physical quantities vary between these bases and thus become unreliable which may lead to inadequate suggestions to improve the magnetic materials.  This nonequivalence, inconsistency, and difficulty in choosing the right basis for the secular approximation urges for a new and more powerful formula of the secular master equation which must be insensitive to the basis selection.

Starting from the general equation governing the relaxation process in the described system, Redfield non-secular master equation, in this work we develop an unprecedented secular form for it. Having this secular form, we apply it to the localized and eigenstates basis to emphasize its difference in these two bases. The paper is thus organized into 5 sections. In particular, we devote Section II for derivation of the secular form of the non-secular master equation regardless of the basis. A general formula for the phonon-induced quantum tunneling rate and correction to transition rate between two energy level are also introduced in this section. Equipped with these, Section III concentrates on the detailed form of the developed ``secular'' non-secular master equation in the localized basis and eigenstates. Last but not least, implications and discussions are given in the last section. 

\section{Secular form of non-secular master equation}

Relaxation of a spin system $S$ with Hamiltonian $\hmt$ in a weakly interaction with a thermal bath can be described by the Redfield master equation \citep{Blum1996,Garanin2011}: 
\begin{equation}
\frac{d\rho}{dt}=-i\left[\hmt,\rho\right]+\hat{R}\rho,\label{eq:Redfield DOE}
\end{equation}
where $\hat{R}$ is the Redfield super-operator. In a specific basis $\left\{ \ket{m}\right\} $, the above equation becomes 
\begin{gather}
\frac{d\rho_{mn}}{dt}=\sum_{k,l}R'_{mn,kl}\rho_{kl},\label{eq:Redfield DME}\\
R_{mn,kl}^{\prime}\equiv R_{mn,kl}+i\left(\delta_{mk}\hmt_{ln}-\hmt_{mk}\delta_{ln}\right).
\end{gather}
Here $R_{mm,mm}=-\sum_{n\ne m}R_{nn,mm}$ is the rate of population loss from state $\ket{m}$ to all other $\ket{n}\ne\ket{m}$; $R_{mm,nn}\equiv\Gamma_{mn}\,\forall n\ne m$ is the population transition rate from $\ket{n}$ to $\ket{m}$; $R_{mn,mn}\equiv-\gamma_{mn}$ is the dephasing rate of $\rho_{mn}$; and other $R_{mn,kl}$ represents the coherence transfer rate at which the density matrix element $\rho_{kl}$ acts upon the change of the element $\rho_{mn}$.

Evidently, equation Eq. \eqref{eq:Redfield DME} is a linear homogeneous system of differential equations with constant coefficients. Consequently, its solution is a linear combination of exponentials $e^{-\lambda_{i}t}$, $i=1\ldots\left(2s+1\right)^{2},$ where $\lambda_{i}$ are eigenvalue of the matrix $R_{mn,kl}^{\prime}$. Since this system of equation describes the relaxation of a physical system, there exists at least one zero eigenvalue corresponding to the equilibrium. Meanwhile, as a majority of those relaxations modes of large eigenvalues will be fast damped in a short time, it is the slow ones (one with small eigenvalue) that decide the total relaxation time of the spin system.

The general solution of Eq. \eqref{eq:Redfield DME} is thus of the following form:
\begin{equation}
\rho_{mn}=\sum_{\mu}\rho_{mn}^{\left(\mu\right)}=\sum_{\mu}A_{mn}^{\left(\mu\right)}e^{-\lambda_{\mu}t}.\label{eq:rho_mn form}
\end{equation}

Redfield equation, Eq. \eqref{eq:Redfield DME}, then becomes: 
\begin{equation}
\sum_{\mu}\left(\frac{d\rho_{mn}^{\left(\mu\right)}}{dt}-\sum_{k,l}R'_{mn,kl}\rho_{kl}^{\left(\mu\right)}\right)=0.
\end{equation}

This equation should be satisfied at any time moment, it is thus equivalent to 
\begin{equation}
\frac{d\rho_{mn}^{\left(\mu\right)}}{dt}=\sum_{k,l}R'_{mn,kl}\rho_{kl}^{\left(\mu\right)}
\end{equation}

Insert Eq. \eqref{eq:rho_mn form} into the above, then solve for $A_{mn}^{\left(\mu\right)}$, we obtain 
\begin{align}
A_{mn}^{\left(\mu\right)} & =\frac{\sum_{\left(kl\right)\notin\left\{ mn,nm\right\} }\left[R'_{mn,nm}R'_{nm,kl}-R'_{mn,kl}\left(R_{nm,nm}^{\prime}+\lambda_{\mu}\right)\right]A_{kl}^{\left(\mu\right)}}{\left(R_{mn,mn}^{\prime}+\lambda_{\mu}\right)\left(R_{nm,nm}^{\prime}+\lambda_{\mu}\right)-\left|R_{mn,nm}^{\prime}\right|^{2}},\,m\ne n,\label{eq:A_mn}
\end{align}
and 
\begin{equation}
A_{mm}^{\left(\mu\right)}=-\frac{\sum_{kl\ne mm}R_{mm,kl}^{\prime}A_{kl}^{\left(\mu\right)}}{\lambda_{\mu}+R_{mm,mm}^{\prime}},\label{eq:A_mm}
\end{equation}
where the term $R_{mn,mn}^{\prime}$ is given by: 
\begin{align}
R_{mn,mn}^{\prime} & =R_{mn,mn}-i\omega_{mn},\\
\omega_{mn} & \equiv\hmt_{mm}-\hmt_{nn}.
\end{align}

In a basis where $\ket{m}$ and $\ket{n}$ mainly reside in the subspace of different $m^{\mathrm{th}}$ and $n^{\mathrm{th}}$ doublet/singlet, $\omega_{mn}$ is approximately the energy gap between these two doublets/singlets and effectively much larger than any $R_{mn,kl}$. The coefficients $A_{mn}^{\left(\mu\right)}$ $\left(m\ne n\right)$ are thus negligible. The only considerable components are $A_{mm}$ and $A_{mm'}$ where $\ket{m}$ and $\ket{m'}$ mainly resides in the same $m^{\mathrm{th}}$ doublet's subspace. This allows us to conclude that provided the above condition of the basis ($\ket{m}$ mainly resides in the subspace of $m^{\mathrm{th}}$ doublet/singlet), the semi-secular approximation is always sufficient in finding the dynamics of the density matrix, i.e. we can approximate: 
\begin{align}
\frac{d\rho_{mm}^{\left(\mu\right)}}{dt} & \approx\sum_{n}\left(R_{mm,nn}^{\prime}\rho_{nn}^{\left(\mu\right)}+R_{mm,nn'}^{\prime}\rho_{nn'}^{\left(\mu\right)}\right),\label{eq:Drho_mm}\\
\frac{d\rho_{mm'}^{\left(\mu\right)}}{dt} & \approx\sum_{n}\left(R_{mm',nn}^{\prime}\rho_{nn}^{\left(\mu\right)}+R_{mm',nn'}^{\prime}\rho_{nn'}^{\left(\mu\right)}\right).\label{eq:Drho_mm'}
\end{align}

These form a closed semi-secular system of equations. In order to have a closed ``secular'' system of equations of the diagonal or off-diagonal density matrix elements only, we further simplify $A_{mm'}^{\left(\mu\right)}$ and $A_{mm}^{\left(\mu\right)}$ using the semi-secular approximation version of Eq. \eqref{eq:A_mn} and \eqref{eq:A_mm} to obtain 
\begin{gather}
A_{mm'}^{\left(\mu\right)}=C_{mm'}^{\left(\mu\right)}\left(A_{mm}^{\left(\mu\right)}-A_{m'm'}^{\left(\mu\right)}\right)+\sum_{k}D_{mm',kk}^{\left(\mu\right)}A_{kk}^{\left(\mu\right)}+\sum_{k\ne m,m'}D_{mm',kk'}^{\left(\mu\right)}A_{kk'}^{\left(\mu\right)},\label{eq:Amm'}\\
A_{mm}^{\left(\mu\right)}=\sum_{k\ne m}D_{mm,kk}^{\left(\mu\right)}A_{kk}^{\left(\mu\right)}+\sum_{k}D_{mm,kk'}^{\left(\mu\right)}A_{kk'}^{\left(\mu\right)},\label{eq:Amm}
\end{gather}
where 
\begin{gather}
C_{mm'}^{\left(\mu\right)}\equiv-i\frac{R_{mm',m'm}\hmt_{m'm}+\hmt_{mm'}\left(\lambda_{\mu}-\gamma_{m'm}+i\omega_{mm'}\right)}{\left(\lambda_{\mu}-\gamma_{m'm}+i\omega_{mm'}\right)\left(\lambda_{\mu}-\gamma_{mm'}-i\omega_{mm'}\right)-\left|R_{mm',m'm}\right|^{2}},\label{eq:Cmm'}\\
D_{mm',kl}^{\left(\mu\right)}\equiv\frac{R_{mm',m'm}R_{m'm,kl}-R_{mm',kl}\left(\lambda_{\mu}-\gamma_{m'm}+i\omega_{mm'}\right)}{\left(\lambda_{\mu}-\gamma_{m'm}+i\omega_{mm'}\right)\left(\lambda_{\mu}-\gamma_{mm'}-i\omega_{mm'}\right)-\left|R_{mm',m'm}\right|^{2}},\\
D_{mm,kk}^{\left(\mu\right)}\equiv\frac{R_{mm,kk}}{\gamma_{mm}-\lambda_{\mu}},\,D_{mm,kk'}^{\left(\mu\right)}\equiv\frac{R_{mm,kk'}+i\left(\delta_{mk}\hmt_{k'm}-\hmt_{mk}\delta_{k'm}\right)}{\gamma_{mm}-\lambda_{\mu}},\label{eq:Dmmkk}
\end{gather}

Eqs. \eqref{eq:Amm'} and \eqref{eq:Amm} are self-consistent. Hence, we can draw similar expressions for $A_{kk'}^{\left(\mu\right)}$ and $A_{kk}^{\left(\mu\right)}$and insert them back to have 
\begin{gather}
A_{mm'}^{\left(\mu\right)}=\sum_{k}\left[F_{mm',kk'}^{\left(\mu\right)}\left(A_{kk}^{\left(\mu\right)}-A_{k'k'}^{\left(\mu\right)}\right)+G_{mm',kk}^{\left(\mu\right)}A_{kk}^{\left(\mu\right)}\right],\label{eq:Amm' final}\\
A_{mm}^{\left(\mu\right)}=\sum_{k}G_{mm,kk'}^{\left(\mu\right)}A_{kk'}^{\left(\mu\right)},\label{eq:Amm final}
\end{gather}
where 
\begin{gather}
F_{mm',kk'}^{\left(\mu\right)}\equiv C_{mm'}^{\left(\mu\right)}\delta_{mk}+\sum_{l\ne m,m'}D_{mm',ll'}^{\left(\mu\right)}C_{ll'}^{\left(\mu\right)}\delta_{lk}+\sum_{l\ne m,m'}\sum_{p\ne l,l'}D_{mm',ll'}^{\left(\mu\right)}D_{ll',pp'}^{\left(\mu\right)}C_{pp'}^{\left(\mu\right)}\delta_{pk}+\ldots,\\
G_{mm',kk}^{\left(\mu\right)}\equiv D_{mm',kk}^{\left(\mu\right)}+\sum_{l\ne m,m'}D_{mm',ll'}^{\left(\mu\right)}D_{ll',kk}^{\left(\mu\right)}+\sum_{l\ne m,m'}\sum_{p\ne l,l'}D_{mm',ll'}^{\left(\mu\right)}D_{ll',pp'}^{\left(\mu\right)}D_{pp',kk}^{\left(\mu\right)}+\ldots,\\
G_{mm,kk'}^{\left(\mu\right)}\equiv D_{mm,kk'}^{\left(\mu\right)}+\sum_{l\ne m}D_{mm,ll}^{\left(\mu\right)}D_{ll,kk'}^{\left(\mu\right)}+\sum_{p\ne m}\sum_{l\ne p}D_{mm,pp}^{\left(\mu\right)}D_{pp,ll}^{\left(\mu\right)}D_{ll,kk'}^{\left(\mu\right)}+\ldots.
\end{gather}
It is worth noting that in the case $\lambda_{\mu}\in\Re$ then $C_{m'm}^{\left(\mu\right)}=-C_{mm'}^{*\left(\mu\right)}$ and $D_{m'm,lk}^{\left(\mu\right)}=D_{mm',kl}^{*\left(\mu\right)}$, accordingly $G_{m'm,kk}^{\left(\mu\right)}=G_{mm',kk}^{*\left(\mu\right)}$ and $F_{m'm,k'k}^{\left(\mu\right)}=-F_{mm',kk'}^{*\left(\mu\right)}$.

Eqs . \eqref{eq:Drho_mm} and \eqref{eq:Drho_mm'} becomes
\begin{align}
\frac{d\rho_{mm}^{\left(\mu\right)}}{dt} & =\sum_{k^{\mathrm{th}}}\Gamma_{m}^{k\left(\mu\right)}\left(\rho_{kk}^{\left(\mu\right)}-\rho_{k'k'}^{\left(\mu\right)}\right)+\sum_{k}\left[R_{mm,kk}+R_{mm,kk}^{\mathrm{\left(corr,\mu\right)}}\right]\rho_{kk}^{\left(\mu\right)},\label{eq:varLambda_mm}\\
\frac{d\rho_{mm'}^{\left(\mu\right)}}{dt} & =\sum_{k}\left(R_{mm',kk'}^{\prime}+R_{mm',kk'}^{\prime\left(\mathrm{corr},\mu\right)}\right)\rho_{kk'}^{\left(\mu\right)},\label{eq:varLambda_mm'}
\end{align}
where 
\begin{gather}
\Gamma_{m}^{k\left(\mu\right)}\equiv i\left[\hmt_{m'm}\left(F_{mm',kk'}^{\left(\mu\right)}-F_{mm',k'k}^{\left(\mu\right)}\right)-\hmt_{mm'}\left(F_{m'm,kk'}^{\left(\mu\right)}-F_{m'm,k'k}^{\left(\mu\right)}\right)\right],\label{eq:Gamma_tunnel_km}\\
R_{mm,kk}^{\mathrm{\left(corr,\mu\right)}}\equiv\Gamma_{mk}^{\mathrm{\left(corr,\mu\right)}}\equiv i\left(\hmt_{m'm}G_{mm',kk}^{\left(\mu\right)}-\hmt_{mm'}G_{m'm,kk}^{\left(\mu\right)}\right)+\sum_{n}R_{mm,nn'}\left(G_{nn',kk}^{\left(\mu\right)}+F_{nn',kk'}^{\left(\mu\right)}-F_{nn',k'k}^{\left(\mu\right)}\right),\\
R_{mm',kk'}^{\prime\left(\mathrm{corr},\mu\right)}\equiv\sum_{n}R_{mm',nn}G_{nn,kk'}^{\left(\mu\right)}+i\hmt_{mm'}\left(G_{mm,kk'}^{\left(\mu\right)}-G_{m'm',kk'}^{\left(\mu\right)}\right).
\end{gather}

Since $R_{mm,mm}=-\sum_{k\ne m}R_{kk,mm}$ and $R_{mm,mm}^{\mathrm{\left(corr\right)}}=-\sum_{k\ne m}R_{kk,mm}^{\mathrm{\left(corr\right)}}$ (see proof in the Appendix), Eqs. \eqref{eq:varLambda_mm} and \eqref{eq:varLambda_mm'} can be rewritten as: 
\begin{align}
\frac{d\rho_{mm}^{\left(\mu\right)}}{dt} & =\sum_{k^{\mathrm{th}}}\Gamma_{m}^{k\left(\mu\right)}\left(\rho_{kk}^{\left(\mu\right)}-\rho_{k'k'}^{\left(\mu\right)}\right)+\sum_{k\ne m}\left[\left(\Gamma_{mk}+\Gamma_{mk}^{\mathrm{\left(corr,\mu\right)}}\right)\rho_{kk}^{\left(\mu\right)}-\left(\Gamma_{km}+\Gamma_{km}^{\mathrm{\left(corr,\mu\right)}}\right)\rho_{mm}^{\left(\mu\right)}\right].\label{eq:rho_mm_final}\\
\frac{d\rho_{mm'}^{\left(\mu\right)}}{dt} & =\sum_{k}\left(R_{mm',kk'}^{\prime}+R_{mm',kk'}^{\prime\left(\mathrm{corr},\mu\right)}\right)\rho_{kk'}^{\left(\mu\right)}.\label{eq:rho_mm'_final}
\end{align}
These are the key equations of the work. As can be seen, the original non-secular density matrix equation now reduces to a closed form of two ``secular'' master equation, one contains only the diagonal density matrix elements and another one contains the off-diagonal density matrix elements. 

From Eqs. (\ref{eq:Gamma_tunnel_km}-\ref{eq:rho_mm'_final}), it is clear to see that $\Gamma_{m}^{k\left(\mu\right)}=-\Gamma_{m}^{k'\left(\mu\right)}=-\Gamma_{m'}^{k\left(\mu\right)}=\Gamma_{m'}^{k'\left(\mu\right)}$plays as the transition rate of population difference from the doublet $k^{\mathrm{th}}$ (representative by $k$ or $k'$), to state $\ket{m}$. Interestingly, $\Gamma_{m}^{m'\left(\mu\right)}=\Gamma_{m'}^{m\left(\mu\right)}$ can be considered as the \emph{general }quantum tunneling rate between two states $\ket{m}$ and $\ket{m'}$ corresponding to the relaxation mode $\lambda_{\mu}$. Meanwhile, $\Gamma_{km}^{\mathrm{\left(corr,\mu\right)}}\,\forall k\ne m$ can be considered as the correction to the transition rate $\Gamma_{km}$ corresponding to the relaxation mode $\lambda_{\mu}$, which results from the semi-secular/non-secular approximation where the time-dependence of off-diagonal density matrix elements is taken into account the time dependence of the diagonal term.  

The series $F_{mm',kk'}^{\left(\mu\right)}$, $G_{mm',kk}^{\left(\mu\right)}$, and $G_{mm,kk'}^{\left(\mu\right)}$ are clearly infinite. However, since $A_{mm'}^{\left(\mu\right)}$ and $A_{mm}^{\left(\mu\right)}$ $\forall m$ are finite, these series must be convergent. We hence can take zeroth-order approximation of them which are $F_{mm',kk'}^{\left(\mu\right)}\approx C_{mm'}^{\left(\mu\right)}\delta_{mk}$, $G_{mm',kk}^{\left(\mu\right)}\approx D_{mm',kk}^{\left(\mu\right)}$, and $G_{mm,kk'}^{\left(\mu\right)}=D_{mm,kk'}^{\left(\mu\right)}$, to find out the zeroth-order approximations of $d\rho_{mm}^{\left(\mu\right)}/dt$ and $d\rho_{mm'}^{\left(\mu\right)}/dt$: 
\begin{gather}
\frac{d\rho_{mm}^{\left(\mu\right)}}{dt}=\Gamma_{m}^{m'\left(\mu,0\right)}\left(\rho_{m'm'}^{\left(\mu\right)}-\rho_{mm}^{\left(\mu\right)}\right)+\sum_{k\ne m}\left[\left(\Gamma_{mk}+\Gamma_{mk}^{\mathrm{\left(corr,\mu,0\right)}}\right)\rho_{kk}^{\left(\mu\right)}-\left(\Gamma_{km}+\Gamma_{km}^{\mathrm{\left(corr,\mu,0\right)}}\right)\rho_{mm}\right],\label{eq:rho_mm_final_0th}\\
\frac{d\rho_{mm'}^{\left(\mu\right)}}{dt}=\sum_{k}\left(R_{mm',kk'}^{\prime}+R_{mm',kk'}^{\prime\left(\mathrm{corr},\mu,0\right)}\right)\rho_{kk'}^{\left(\mu\right)},\label{eq:rho_mm'_final_0th}
\end{gather}
 where 
\begin{gather}
\Gamma_{m}^{m'\left(\mu,0\right)}=-i\left(\hmt_{m'm}C_{mm'}^{\left(\mu\right)}+\hmt_{mm'}C_{m'm}^{\left(\mu\right)}\right),\label{eq:Gamma_tunnel_0}\\
\Gamma_{mk}^{\mathrm{\left(corr,\mu,0\right)}}=i\left(\hmt_{m'm}D_{mm',kk}^{\left(\mu\right)}-\hmt_{mm'}D_{m'm,kk}^{\left(\mu\right)}\right)+\left(R_{mm,kk'}C_{kk'}^{\left(\mu\right)}-R_{mm,k'k}C_{k'k}^{\left(\mu\right)}\right)+\sum_{n}R_{mm,nn'}D_{nn',kk}^{\left(\mu\right)},\,\forall k\ne m,\label{eq:Gamma_corr_0}\\
R_{mm',kk'}^{\prime\left(\mathrm{corr},\mu,0\right)}=\sum_{n}R_{mm',nn}D_{nn,kk'}^{\left(\mu\right)}+i\hmt_{mm'}\left(D_{mm,kk'}^{\left(\mu\right)}-D_{m'm',kk'}^{\left(\mu\right)}\right).\label{eq:R_mmkk'_corr_0}
\end{gather}
It is obvious from above that only the quantum tunneling rate between $\ket{m}$ and $\ket{m'}$ enters the zeroth-order approximation of the equation of state of $\rho_{mm}$. This underscores the important role of the quantum tunneling within each doublet to its relaxation compared to contributions from the population difference of other doublets.

It is important to note that since all $\vt{\rho}^{\left(\mu\right)}$ and $\lambda_{\mu}$ are mathematically equivalent, the subscript $\mu$ can be dropped from all above equations. 

\section{\textquotedblleft Secular\textquotedblright{} non-secular master equation in localized \& eigenstates basis}

Up to now, a vast majority of the magnetization relaxation studies uses the localized and eigenstates basis. Hence, in the following we show the detailed form of the quantum tunneling rate, the correction to other transition rates, and the secular form in these bases by consecutively apply our developed formulas in the previous section. For simplicity, only the zeroth-order approximation of $d\rho_{mm}/dt$ and $d\rho_{mm'}/dt$ are retained.

\subsection{Localized basis}

Hamiltonian of a spin system corresponding to $m^{\mathrm{th}}$ doublet subspace in the presence of a magnetic field takes the following form in the localized basis \citep{Gatteschi2006,Leuenberger2000,Ho2018,Ho2022a}: 
\begin{equation}
\hmt_{\left\{ \ket{m},\ket{m'}\right\} }=\frac{W_{mm'}}{2}\left(\ket{m}\bra{m}-\ket{m'}\bra{m'}\right)+\frac{\Delta_{mm'}}{2}\left(\ket{m}\bra{m'}+\ket{m'}\bra{m}\right),
\end{equation}
where $W_{mm'}$ is the energy bias between $\ket{m}$ and $\ket{m'}$ and $\Delta_{mm'}$ is the tunneling splitting of $m^{\mathrm{th}}$ doublet. Without loss of generality, these two quantities can be taken to be real numbers. With this Hamiltonian, we have: 
\begin{gather}
C_{mm'}=\frac{i\Delta_{mm'}}{2}\frac{\left(\gamma_{mm'}-\lambda\right)-iW_{mm'}-R_{mm',m'm}}{\left(\gamma_{mm'}-\lambda\right)^{2}+W_{mm'}^{2}-\left|R_{mm',m'm}\right|^{2}},\\
D_{mm',kl}=\frac{R_{mm',m'm}R_{m'm,kl}+R_{mm',kl}\left[\gamma_{mm'}-\lambda-iW_{mm'}\right]}{\left(\gamma_{mm'}-\lambda\right)^{2}+W_{mm'}^{2}-\left|R_{mm',m'm}\right|^{2}},\\
D_{mm,kk}\equiv\frac{R_{mm,kk}}{\gamma_{mm}-\lambda_{\mu}},\,D_{mm,kk'}\equiv\frac{R_{mm,kk'}+i\Delta_{mm'}\left(\delta_{mk}-\delta_{mk'}\right)}{\gamma_{mm}-\lambda_{\mu}},
\end{gather}
where $\gamma_{mm'}\equiv-R_{mm',mm'}$ is real-valued, which is due to 1) the property of the Redfield super-operator matrix element $R_{mm',mm'}=R_{m'm,m'm}^{*}$; 2) the time-reversal symmetry between two localized states which allows us to consider $\ket{m}$ and $\ket{m'}$ as ``up'' and ``down'' state of the $m^{\mathrm{th}}$ doublet \citep{Ho2022a} and accordingly $R_{mm',mm'}=R_{m'm,m'm}$. Substituting these into Eqs. \eqref{eq:Gamma_tunnel_0} and \eqref{eq:Gamma_corr_0} results in the quantum tunneling rate within the $m^{\mathrm{th}}$ doublet:
\begin{align}
\Gamma_{m}^{m'} & =\frac{\Delta_{mm'}^{2}}{2}\frac{\gamma_{mm'}-\lambda-\left(R_{mm',m'm}+R_{m'm,mm'}\right)/2}{\left(\gamma_{mm'}-\lambda\right)^{2}+W_{mm'}^{2}-\left|R_{mm',m'm}\right|^{2}},\label{eq:Gamma_tunnel_0-localized}
\end{align}

Magnetization relaxation of the spin is determined by slow modes. Hence, we can ignore large eigenvalues $\lambda_{\mu}$ and make further approximations. In particular, for excited doublet over the whole temperature range, it is typical that $\gamma_{mm'}\gg\lambda$ (for slow modes) and $\gamma_{mm'}\gg\left|R_{mm',m'm}\right|,\left(R_{mm',m'm}+R_{m'm,mm'}\right)$, a simpler version of the quantum tunneling rate within the excited doublets $\Gamma_{m}^{m'}$, Eq. \eqref{eq:Gamma_tunnel_0-localized}, can thus be used:
\begin{align}
\Gamma_{m}^{m'} & =\frac{\Delta_{mm'}^{2}}{2}\frac{\gamma_{mm'}^{\prime}}{\gamma_{mm'}^{\prime2}+W_{mm'}^{2}},\label{eq:Incoherent tunneling rate}
\end{align}
which is basically similar to the well-known incoherent tunneling rate \citep{Garanin1997,Leuenberger2000}. 

For the ground doublet at low temperature, those above conditions for $\gamma_{mm'}$ are hardly fulfilled. A direct application of Eq. \eqref{eq:Incoherent tunneling rate} may thus give inaccurate results. Accordingly a full treatment of the quantum tunneling rate $\Gamma_{m}^{m'}$ is essential. Additionally, in this temperature range, there are many co-existing slow relaxation modes \citep{Ho2022b,Ho2022c}, accordingly multiple $\Gamma_{m}^{m'\left(\mu\right)}$ corresponding to these slow modes are thus needed as well. Consequently, it is inconvenient to deal with this issue of the quantum tunneling of magnetization in the ground doublet within the current framework here. We reserve this in other works \citep{Ho2022b,Ho2022c}. Meanwhile, the validity of the incoherent tunneling rate for the ground doublet at high temperature is certainly subject to the relative magnitude between the dephasing rate $\gamma_{mm'}$ and $\lambda$.

Similarly to the quantum tunneling rate, we have the following expression for $\Gamma_{mk}^{\mathrm{\left(corr\right)}}$:
\begin{align}
\Gamma_{mk}^{\mathrm{\left(corr\right)}} & =\frac{i\Delta_{mm'}}{2}\frac{R_{m'm,kk}\left[R_{mm',m'm}-\left(\gamma_{mm'}-\lambda\right)-iW_{mm'}\right]}{\left(\gamma_{mm'}-\lambda\right)^{2}+W_{mm'}^{2}-\left|R_{mm',m'm}\right|^{2}}\nonumber \\
 & \quad+\frac{i\Delta_{kk'}}{2}\frac{R_{mm,k'k}\left[R_{k'k,kk'}-\left(\gamma_{kk'}-\lambda\right)-iW_{kk'}\right]}{\left(\gamma_{kk'}-\lambda\right)^{2}+W_{kk'}^{2}-\left|R_{kk',k'k}\right|^{2}}\nonumber \\
 & \quad+\sum_{n^{\mathrm{th}}}R_{mm,nn'}\frac{R_{nn',n'n}R_{n'n,kk}+R_{nn',kk}\left[\left(\gamma_{nn'}-\lambda\right)-iW_{nn'}\right]}{\left(\gamma_{nn'}-\lambda\right)^{2}+W_{nn'}^{2}-\left|R_{nn',n'n}\right|^{2}},\nonumber \\
 & \quad+\mathrm{h.c.}\left(\lambda^{*}\rightarrow\lambda\right),\,\forall k\ne m.\label{eq:Gamma_mk_0_corr_localized}
\end{align}

At high temperature we can also approximate $\Gamma_{mk}^{\mathrm{\left(corr\right)}}$ by simply neglecting $\lambda$ in Eq. \eqref{eq:Gamma_mk_0_corr_localized}. In the case where the temperature is sufficiently high that the coherence transfer rates are much smaller than the dephasing rates, i.e. $R_{nn',n'n},R_{n'n,kk}\ll\left|\gamma_{nn'}\right|\,\forall n,k$, $\Gamma_{mk}^{\mathrm{\left(corr\right)}}$ becomes 
\begin{equation}
\Gamma_{mk}^{\mathrm{\left(corr\right)}}\approx\frac{i}{2}\left(\frac{\Delta_{mm'}R_{mm',kk}}{\gamma_{mm'}+iW_{mm'}}+\frac{\Delta_{kk'}R_{mm,kk'}}{\gamma_{kk'}+iW_{kk'}}\right)+\mathrm{h.c.}\,\forall k\ne m.\label{eq:Gamma_mk_0_corr_localized-1-1}
\end{equation}

As can be seen, this correction may not be small compared to $\Gamma_{mk}$ even at high temperature if the tunneling splittings are large and near resonance. Accordingly, the correction to the transition rate $\Gamma_{mk}^{\mathrm{\left(corr\right)}}$, besides the quantum tunneling rate $\Gamma_{m}^{m'}$, should be taken into account when magnetization relaxation is studied in localized basis. 

In short, the ``secular'' non-secular master equation in localized basis is not so much different from the general one in previous section. In particular, 
\begin{align*}
\frac{d\rho_{mm}}{dt} & =\Gamma_{m}^{m'}\left(\rho_{m'm'}-\rho_{mm}\right)+\sum_{k\ne m}\left[\left(\Gamma_{mk}+\Gamma_{mk}^{\mathrm{\left(corr\right)}}\right)\rho_{kk}-\left(\Gamma_{km}+\Gamma_{km}^{\mathrm{\left(corr\right)}}\right)\rho_{mm}\right],
\end{align*}
where all transition rates and their correction are given above. As a limiting case when the system is out of resonance and  accordingly both quantum tunneling rate and the corrections to the transition rates becomes negligible, our equation for diagonal density matrix elements reduces to the non-tunneling secular master equation as expected.

For off-diagonal elements of the density matrix, we have the following correction term: 
\begin{align*}
R_{mm',kk'}^{\prime\left(\mathrm{corr}\right)} & =\frac{i}{2}\frac{\Delta_{mm'}}{\gamma_{mm}-\lambda}\left(R_{mm,kk'}-R_{m'm',kk'}\right)+\frac{i}{2}\frac{\Delta_{kk'}}{\gamma_{kk}-\lambda}\left(R_{mm',kk}-R_{mm',k'k'}\right)\\
 & \qquad\qquad+\frac{\Delta_{mm'}^{2}}{\gamma_{mm}-\lambda}\left(\delta_{m'k}-\delta_{mk}\right)+\sum_{n}\frac{R_{mm',nn}R_{nn,kk'}}{\gamma_{nn}-\lambda},
\end{align*}
and accordingly, 
\begin{multline*}
\frac{d\rho_{mm'}}{dt}=-\left(\gamma_{mm'}+iW_{mm'}\right)\rho_{mm'}-\frac{1}{2}\frac{\Delta_{mm'}^{2}}{\gamma_{mm}-\lambda}\left(\rho_{mm'}-\rho_{m'm}\right)\\
+\frac{i}{2}\frac{\Delta_{mm'}}{\gamma_{mm}-\lambda}\left(R_{mm',mm}-R_{mm',m'm'}+R_{mm,mm'}-R_{m'm',mm'}\right)\left(\rho_{mm'}-\rho_{m'm}\right)\\
+\sum_{k\ne m}R_{mm',kk'}\rho_{kk'}+\sum_{n,k}\frac{R_{mm',nn}R_{nn,kk'}}{\gamma_{nn}-\lambda}+\frac{i}{2}\frac{\Delta_{mm'}}{\gamma_{mm}-\lambda}\sum_{k\ne m,m'}\left(R_{mm,kk'}-R_{m'm',kk'}\right)\rho_{kk'}.
\end{multline*}
At high temperature, we can ignore $\lambda$ and all coherence transfer rates, which results in: 
\begin{align*}
\frac{d\rho_{mm'}}{dt} & =-\left(\gamma_{mm'}+iW_{mm'}\right)\rho_{mm'}-\frac{1}{2}\frac{\Delta_{mm'}^{2}}{\gamma_{mm}}\left(\rho_{mm'}-\rho_{m'm}\right).
\end{align*}

\subsection{Eigenstates basis}

In eigenstates basis, the Hamiltonian of the spin system is diagonal: 
\begin{equation}
\hmt=\sum_{\alpha^{\mathrm{th}}}\left(\varepsilon_{\alpha}\ket{\alpha}\bra{\alpha}+\varepsilon_{\alpha'}\ket{\alpha'}\bra{\alpha'}\right),
\end{equation}
where $\ket{\alpha}$ and $\ket{\alpha'}$ belong to the same doublet $\alpha^{\mathrm{th}}$. In this basis, $\gamma_{\alpha\alpha'}=\gamma_{\alpha'\alpha}$ and both are also real \citep{Blum1996,Garanin2011}. 

From Eqs. (\ref{eq:Cmm'}-\ref{eq:Dmmkk}), it is easy to see that 
\begin{gather}
C_{\alpha\alpha'}=0,\,\,\,D_{\alpha\alpha,\beta\beta'}=\frac{R_{\alpha\alpha,\beta\beta'}}{\gamma_{\alpha\alpha}-\lambda},\\
D_{\alpha\alpha',\beta\gamma}=\frac{R_{\alpha\alpha',\alpha'\alpha}R_{\alpha'\alpha,\beta\gamma}+R_{\alpha\alpha',\beta\gamma}\left(\gamma_{\alpha\alpha'}-\lambda-i\omega_{\alpha\alpha'}\right)}{\left(\gamma_{\alpha\alpha'}-\lambda\right)^{2}+\omega_{\alpha\alpha'}^{2}-\left|R_{\alpha\alpha',\alpha'\alpha}\right|^{2}}.
\end{gather}

Consequently, we have
\begin{gather}
\Gamma_{\alpha}^{\beta}=0\,\,\,\forall\beta,\\
\Gamma_{\alpha\beta}^{\mathrm{\left(corr\right)}}=\sum_{\delta}R_{\alpha\alpha,\delta\delta'}\frac{R_{\delta\delta',\delta'\delta}R_{\delta'\delta,\beta\beta}+R_{\delta\delta',\beta\beta}\left(\gamma_{\delta\delta'}-\lambda-i\omega_{\delta\delta'}\right)}{\left(\gamma_{\delta\delta'}-\lambda\right)^{2}+\omega_{\delta\delta'}^{2}-\left|R_{\alpha\alpha',\alpha'\alpha}\right|^{2}},\,\,\,\forall\beta\ne\alpha,\\
R_{\alpha\alpha',\beta\beta'}^{\prime\left(\mathrm{corr}\right)}=\sum_{\delta}\frac{R_{\alpha\alpha',\delta\delta}R_{\delta\delta,\beta\beta'}}{\gamma_{\delta\delta}-\lambda},\label{eq:Gamma_ab_corr}
\end{gather}

Interestingly, the phonon-induced quantum tunneling rate in the eigenstates basis is \emph{zero }to any order of approximation of $\Gamma_{\alpha}^{\beta}$. Additionally, at high temperature, $\lambda$ can be ignored from the above to obtain simpler expressions. 

At very low temperature, all the coherence transfer rates entering the expression of $\Gamma_{\alpha\beta}^{\mathrm{\left(corr\right)}}$ is negligible compared to $\left|\omega_{11'}\right|=\sqrt{\Delta_{11'}^{2}+W_{11'}^{2}}\ge\left|\Delta_{11'}\right|$ of the ground doublet. Accordingly, we can neglect $\Gamma_{\alpha\beta}^{\mathrm{\left(corr\right)}}$ and the master equation for diagonal density matrix elements reduces to the secular one. Hence, it is convenient to use the eigenstates basis in this temperature regime. In contrast, if we look at Eq. \eqref{eq:Gamma_mk_0_corr_localized} of $\Gamma_{mk}^{\mathrm{\left(corr\right)}}$in localized basis, the corresponding term entering the denominator is the energy bias $W_{11'}$ instead of $\omega_{11'}$. This energy bias is caused by the magnetic field and may not be small. Accordingly, $\Gamma_{mk}^{\mathrm{\left(corr\right)}}$ in localized basis cannot be ignored at low temperature. This difference in the correction transition rates, and the quantum tunneling rate above, in the same temperature domain underlines the importance of choosing the correct basis for secular master equation in magnetization relaxation problem.

With $\Gamma_{\alpha\beta}^{\mathrm{\left(corr\right)}}$ and $R_{\alpha\alpha',\beta\beta'}^{\prime\left(\mathrm{corr}\right)}$ above, the ``secular'' non-secular master equation in eigenstates basis is then of a simple form: 
\begin{gather}
\frac{d\rho_{\alpha\alpha}}{dt}=\sum_{\beta\ne\alpha}\left[\left(\Gamma_{\alpha\beta}+\Gamma_{\alpha\beta}^{\mathrm{\left(corr\right)}}\right)\rho_{\beta\beta}-\left(\Gamma_{\beta\alpha}+\Gamma_{\beta\alpha}^{\mathrm{\left(corr\right)}}\right)\rho_{\alpha\alpha}\right],\\
\frac{d\rho_{\alpha\alpha'}}{dt}=-i\omega_{\alpha\alpha'}\rho_{\alpha\alpha'}+\sum_{\beta}\left(R_{\alpha\alpha',\beta\beta'}+\sum_{\delta}\frac{R_{\alpha\alpha',\delta\delta}R_{\delta\delta,\beta\beta'}}{\gamma_{\delta\delta}-\lambda}\right)\rho_{\beta\beta'}.
\end{gather}

As can be seen, the ``secular'' non-secular master equation in eigenstates basis reduces to the familiar secular one when correction rates $\Gamma_{\alpha\beta}^{\mathrm{\left(corr\right)}}$ and $R_{\alpha\alpha',\beta\beta'}^{\prime\left(\mathrm{corr}\right)}$ are negligible. From above expressions, this occurs when the splittings $\omega_{\delta\delta'}$ is much larger than all coherence transfer rates. This either results from a large magnetic field (internal or external), which causes large energy biases, or at very low temperature when all coherence transfer rates are minuscule, or the spin system have a low anisotropy and accordingly large tunneling splittings. These deems to be the conditions for using the secular master equation in eigenstates basis at a particular temperature.

\section{Discussions}

Up to now, one may wonder about the use case of our developed secular form since it is self-consistent, i.e. to find the relaxation rate we need to know in advance the relaxation rate. However, we emphasize here that the first purpose of the secular form is to determine the correctness of the secular approximation in addressing the magnetization relaxation problem when it is used in a specific basis. On this aspect, our secular form helps to facilitate different approximations to the Redfield equation based on the information about the operating condition of the system. This includes, but is not limited to, the usual secular approximation in eigenstates/localized states basis. Hence, the exact prior knowledge of relaxation rate is not necessary. Furthermore and similar to any self-consistent method, wherever a specific value of the relaxation rate is needed to deal with the relaxation problem, an arbitrary value can be assigned (typically zero) at the beginning. The developed secular form is then used as a measure to verify the correctness of the solution or to consider the repetition of the solving process using a better initial estimation of the relaxation rate. 

With the advancement of \emph{ab initio} calculation recently, it is not far-fetched to imagine a future where the relaxation rate can be numerically calculated from the crystal structure of the magnetic material \citep{Lunghi2022,Briganti2021a,Garlatti2021a,Lunghi2020a,lunghi2020insights} and the Redfield non-secular master equation. In this scenario, the relaxation rate can be computed \emph{ab initio}. However, to understand the physics and accordingly optimize the molecular/crystal structure for better magnetic performance, the relaxation pathway of the magnetization is indispensable. While solving the non-secular master equation with its inherent complexity can give us a value for the relaxation rate, it is unable to paint a relaxation pathway picture which is necessary to unravel the underlying relaxation mechanism. This picture essentially still relies on the secular approximation and the popular transition rates between states. This is the place where our secular form shines. Indeed, since our secular form inherits the accuracy of the non-secular master equation regardless of the basis and at the same time allows to associate each relaxation path between any two states with a corrected population transition rate, it can fulfill this purpose in a perfect way. 

In summary, the present study has settled the debate over which basis should be chosen when the secular approximation is used. By developing a universal secular form for the Redfield non-secular master equation, the choice of the basis, whether it is localized or eigenstates or some special one, is no longer important. Based on this finding, a general formula of the phonon-induced quantum tunneling rate and correction to other transition rates are also proposed. We found that a proper treatment of the spin relaxation in localized basis must take into account the corrections to both familiar quantum tunneling rate and population transition rates. Meanwhile, in eigenstates basis, our results clearly demonstrated that the phonon-induced quantum tunneling rate in this basis is always equal to zero. Last but not least, the developed secular form is valid in any condition and readily used for any system provided that the Redfield non-secular master equation is relevant. 

\begin{acknowledgments}
L. T. A. H. acknowledge the financial support of the research projects R-143-000-A65-133, A-8000709-00-00, A-8000017-00-00 of the National University of Singapore; and the Flemish Science Foundation (FWO) during his PhD. Calculations were done on the ASPIRE-1 cluster (www.nscc.sg) under the projects 11001278 and 51000267. Computational resources of the HPC-NUS are gratefully acknowledged.
\end{acknowledgments}

\appendix

\section*{Appendix: relaxation matrix elements in an arbitrary basis\label{sec:Some-properties-of}}
\begin{itemize}
\item Relaxation matrix elements $R_{\alpha\beta,\alpha'\beta'}$ in the eigenstates basis can be calculated from the spin-phonon Hamiltonian $V$ \citep{Garanin2011}: 
\begin{align}
R_{\alpha\beta,\alpha'\beta'} & =\frac{\pi}{\hbar Z_{\mathrm{b}}}\sum_{ww'}\left\{ -\sum_{\gamma}e^{-E_{w}/kT}\delta\left(\varepsilon_{\alpha'}-\varepsilon_{\gamma}+E_{w}-E_{w'}\right)V_{\alpha w,\gamma w'}V_{\gamma w',\alpha'w}\delta_{\beta\beta'}\right.\\
 & \quad-\sum_{\gamma}e^{-E_{w}/kT}\delta\left(\varepsilon_{\beta'}-\varepsilon_{\gamma}+E_{w}-E_{w'}\right)V_{\beta'w,\gamma w'}V_{\gamma w',\beta w}\delta_{\alpha\alpha'}\\
 & \left.\quad+e^{-E_{w'}/kT}\left[\delta\left(\varepsilon_{\beta}-\varepsilon_{\beta'}+E_{w}-E_{w'}\right)+\delta\left(\varepsilon_{\alpha}-\varepsilon_{\alpha'}+E_{w}-E_{w'}\right)\right]V_{\alpha w,\alpha'w'}V_{\beta'w',\beta w}\right\} ,
\end{align}
where $\ket{w}$ and $\ket{w'}$ designate the eigenstates of the thermal bath and $Z_{\mathrm{b}}$ is the bath partition function.
\item Relaxation matrix elements $R_{mn,kl}$ in an arbitrary basis $\left\{ \ket{m}\right\} $ are related to the elements $R_{\alpha\beta,\gamma\delta}$ in the eigenstates basis as follows \citep{Garanin2011}: 
\begin{equation}
R_{mn,kl}=\sum_{\alpha,\beta,\gamma,\delta}\braket{m|\alpha}\braket{\beta|n}\braket{\gamma|k}\braket{l|\delta}R_{\alpha\beta,\gamma\delta},
\end{equation}
and vice versa,
\begin{equation}
R_{\alpha\beta,\gamma\delta}=\sum_{m,n,k,l}\braket{\alpha|m}\braket{n|\beta}\braket{k|\gamma}\braket{\delta|l}R_{mn,kl}.
\end{equation}
\item $R_{mn,kl}^{*}=R_{nm,kl}$ -- Proof: this property can be easily derived from the above relation between $R_{mn,pq}$ and $R_{\alpha\beta,\gamma\delta}$, 
\begin{align}
R_{mn,kl}^{*} & =\sum_{\alpha,\beta,\gamma,\delta}\braket{m|\alpha}^{*}\braket{\beta|n}^{*}\braket{\gamma|k}^{*}\braket{l|\delta}R_{\alpha\beta,\gamma\delta}^{*}\nonumber \\
 & =\sum_{\beta,\alpha,\delta,\gamma}\braket{n|\beta}\braket{\alpha|m}\braket{\delta|l}\braket{k|\gamma}R_{\beta\alpha,\delta\gamma}=R_{nm,kl}.
\end{align}
From this, it is straightforward to show that $R_{mn,kl}^{\prime*}=R_{nm,kl}^{\prime}$, $C_{m'm}=-C_{mm'}^{*}$, and $D_{m'm,lk}=D_{mm',kl}^{*}$.
\item $\sum_{m}R_{mm,kl}=0$ -- Proof: from the property $\sum_{\gamma}R_{\alpha\alpha,\gamma\delta}=0$, we have 
\begin{align}
\sum_{m}R_{mm,kl} & =\sum_{m}\sum_{\alpha,\beta,\gamma,\delta}\braket{m|\alpha}\braket{\beta|m}\braket{\gamma|k}\braket{l|\delta}R_{\alpha\beta,\gamma\delta}\nonumber \\
 & =\sum_{\alpha,\beta,\gamma,\delta}\delta_{\alpha\beta}\braket{\gamma|k}\braket{l|\delta}R_{\alpha\beta,\gamma\delta}\nonumber \\
 & =\sum_{\gamma,\delta}\braket{\gamma|k}\braket{l|\delta}\left(\sum_{\alpha}R_{\alpha\alpha,\gamma\delta}\right)=0.
\end{align}
\\
As a result, $R_{mm,mm}=-\sum_{k\ne m}R_{kk,mm}$. 
\item $\sum_{k}R_{kk,mm}^{\mathrm{\left(corr\right)}}=0$ -- Proof: 
\begin{align}
\sum_{k}R_{kk,mm}^{\mathrm{\left(corr\right)}} & =i\sum_{k}\left(\hmt_{k'k}G_{kk',mm}-\hmt_{kk'}G_{k'k,mm}\right)+\sum_{n}\left(\sum_{k}R_{kk,nn'}\right)\left(G_{nn',mm}+F_{nn',mm'}-F_{nn',m'm}\right)\\
 & =i\sum_{k^{\mathrm{th}}}\left[\left(\hmt_{k'k}G_{kk',mm}-\hmt_{kk'}G_{k'k,mm}\right)+\left(\hmt_{kk'}G_{k'k,mm}-\hmt_{k'k}G_{kk',mm}\right)\right]=0
\end{align}
\end{itemize}
\bibliographystyle{apsrev4-2}
\bibliography{reference}

\end{document}